\documentstyle[12pt,psfig]{article}
\def\reference{\parskip 0pt\par\noindent\hangindent 0.5 truecm}

\begin{document}
%
%
\title{At the Vigintennial of the Butcher--Oemler Effect}

\def\gs{\mathrel{\raise0.35ex\hbox{$\scriptstyle >$}\kern-0.6em
\lower0.40ex\hbox{{$\scriptstyle \sim$}}}}
\def\ls{\mathrel{\raise0.35ex\hbox{$\scriptstyle <$}\kern-0.6em
\lower0.40ex\hbox{{$\scriptstyle \sim$}}}}
\def\ergs {{\rm erg} \, {\rm s}^{-1}}
\def\deg {{^{\circ}}}
\def\mnras {{MNRAS}}
\def\apj {ApJ}
\def\apjs {ApJS}
\def\apjl {ApJ}
\def\aj {AJ}
\def\aaps {A\&AS}
\def\araa {ARA\&A}
\def\pasj {PASJ}
\def\nat {Nature}


\author{Kevin A.\ Pimbblet
} 

\date{}
\maketitle

{\center
Department of Physics, University of Queensland, Brisbane, Queensland 4072, Australia\\ \ \\
pimbblet@physics.uq.edu.au\\[3mm]
}

%
\begin{abstract}
In their study of the evolution of galaxies within clusters, Butcher and Oemler discovered evidence for a strong
evolution in star-formation rate with redshift. Later studies confirmed this conclusion and uncovered several
aspects of the effect: photometric, spectroscopic, and morphological. This article reviews a broad sample of these
works and discusses selection effects, biases, and driving mechanisms that might be responsible for the changes in
star-formation rate.

\end{abstract}

{\bf Keywords:} galaxies: evolution --- galaxies: clusters: general --- cosmology: observations

\bigskip

%
%

\section{Introduction}
\label{one} About twenty years ago, Butcher and Oemler published a series of influential papers investigating the
evolution of galaxies within clusters of galaxies (Butcher \& Oemler 1978a, 1978b, 1984, 1985; Butcher, Wells, \&
Oemler 1983). From this series, it is the 1984 paper that describes the Butcher--Oemler (BO hereafter) effect. To
understand the BO effect, it is necessary to examine the observational foundations upon which it is built.

Clusters of galaxies allow the study of a large number of galaxies ($\sim 10^3$) at a common distance and visible
out to a large redshift, which makes them ideal for studying galaxy evolution. They feature predominantly
passively evolving early-type (elliptical and lenticular) galaxies in their core regions (e.g.\ Pimbblet 2001). It
is these galaxies that Visvanathan \& Sandage (1977) first noted exhibit systematically redder integrated colours
than the late-type (spiral and irregular) galaxies in clusters. Further, the early-type galaxies also demonstrate
a tight correlation between their colours and magnitudes: a colour--magnitude relationship (CMR hereafter: e.g.\
Bower, Lucey, and Ellis 1992). By calculating $R_{30}$, the projected clustocentric radius that contains 30 per
cent of the total galaxy population, Butcher \& Oemler (1984) found the excess fraction of galaxies bluer than the
CMR. Specifically, this blue fraction, $f_B$, is defined to be the fraction of galaxies within $R_{30}$ brighter
than $M_V=-20$ whose rest frame $(B-V)$ colour is at least 0.2 magnitudes bluer than the cluster's CMR.

Importantly, they also uncovered evidence for an evolution
in the blue fraction with redshift,
increasing from $f_B \sim 0.03$ at low redshifts ($z \sim 0.05$) to $f_B \sim 0.25$ at larger redshifts ($z \sim
0.5$). It is this evolution that is commonly referred to as the BO effect. The implication of these results is
that the star-formation rate within galaxies has decreased dramatically with epoch irrespective of local
environment. This article presents a historical review of progress into the understanding of the origins of the BO
effect. In Section 2, the confirmation of the BO effect is presented alongside some immediate complicating
factors. Section 3 discusses attempts to tie together several strands of observational evidence (photometric,
spectroscopic, and morphological), whilst in Section 4, possible mechanisms for the effect are discussed. How
selection effects can alter the calculated values of $f_B$ are discussed in more detail in Section 5. Finally,
some concluding remarks are made in Section 6.

\section{Confirmations and Complications}

The original work has been followed up by many other authors (e.g. Couch \& Newell 1984; Rakos \& Schombert 1995)
who show that the cores of $z>0.2$ clusters do contain blue galaxies, complementing the work of other authors
(e.g. Terlevich, Caldwell, \& Bower 2001) who find uniformly red, early-type galaxies populating local nearby
clusters. The natural conclusion to this situation supports the original one: there has been strong, rapid
evolution of clusters of galaxies with redshift.

\begin{figure}
\centerline{\psfig{file=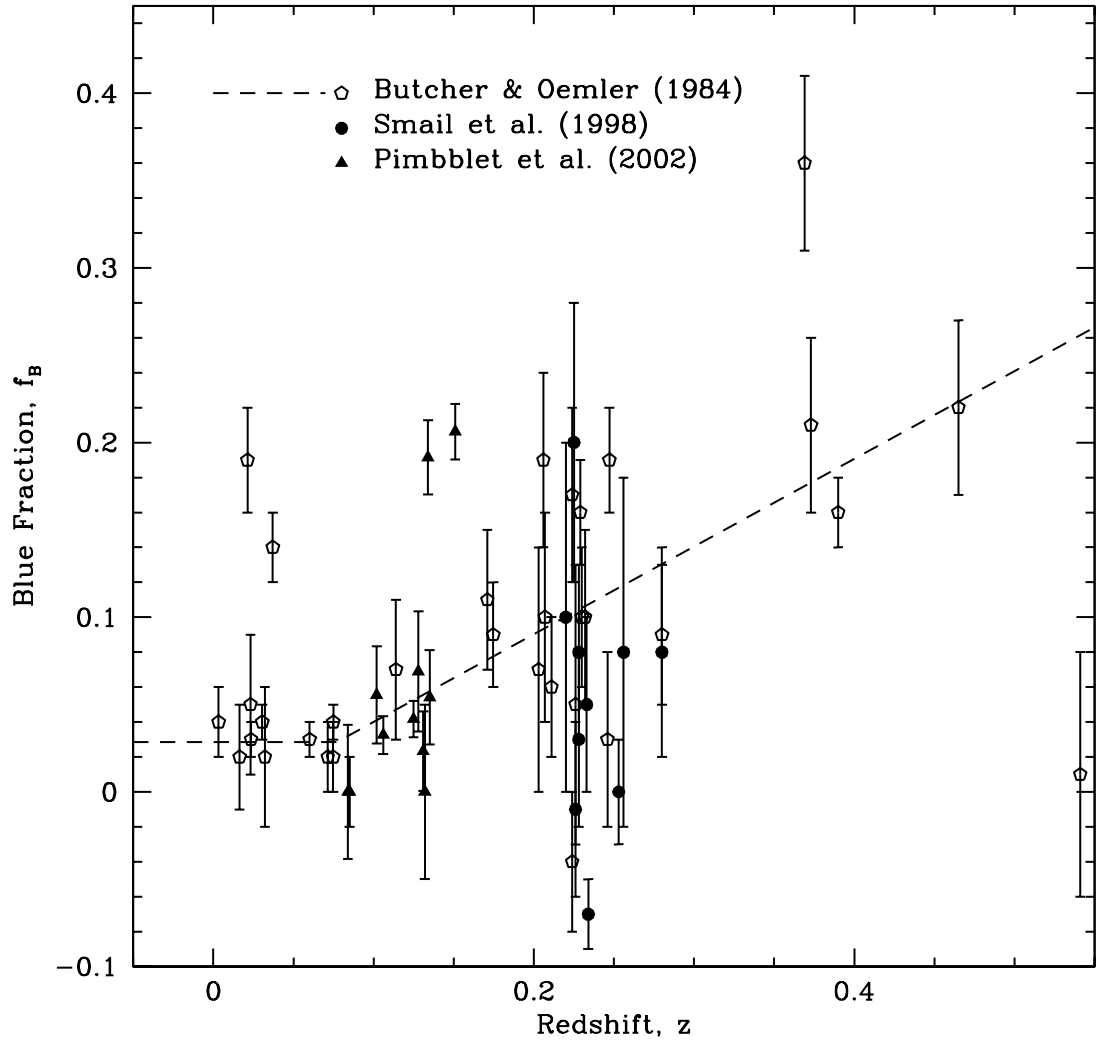,angle=0,width=6.in}}
  \caption[The Butcher--Oemler effect]{\small{The Butcher-Oemler
effect, showing an evolution in the blue fraction with redshift (dashed line). Also plotted are the points from
Smail et al.\ (1998) (filled circles) and Pimbblet et al.\ (2002) (filled triangles) for comparison. }}
  \label{fig:bo84}
\end{figure}

More recently, Smail et al.\ (1998) examined a sample of 10 rich clusters of galaxies in the redshift range
$0.22<z<0.28$ and found that the blue fraction (median $f_B \sim 0.05$) is lower than the values found by Butcher
\& Oemler (1984; see Figure \ref{fig:bo84}). The individual measurements, however, show a large scatter (see also
La Barbera et al.\ 2003). There is also found to be no significant correlation between $f_B$ and global cluster
parameters such as X-ray morphology or cluster concentration. Other studies (e.g.\ Margoniner et al.\ 2001;
Pimbblet et al.\ 2002; Figure \ref{fig:bo84}) display a similarly large scatter in $f_B$.

%

\begin{figure}
\centerline{\psfig{file=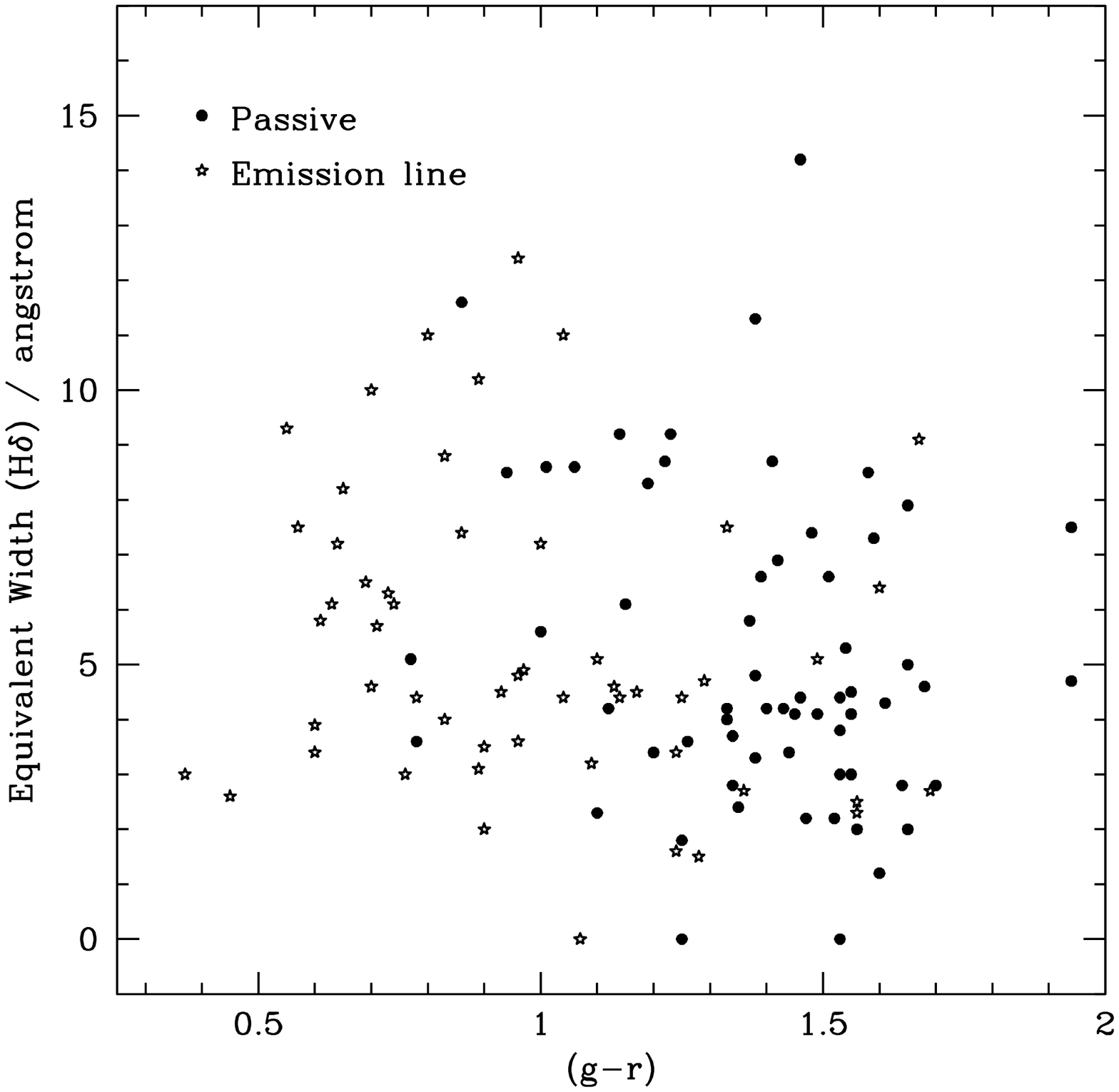,angle=0,width=6.in}}
  \caption[Data from Poggianti et al.\ (1999)]{\small{Distribution
of cluster galaxies (mean redshift $z\sim 0.44$) adapted from data presented by Poggianti et al.\ (1999) and
Dressler et al.\ (1999) on a $(g-r)$ colour--H$\delta$ plane.  Passive galaxies are marked with filled circles and
[OII] emission line galaxies with open stars.  Many of the passive galaxies possess a strong H$\delta$ equivalent
width, indicative of recent star-formation.}}
  \label{fig:p99}
\end{figure}

To test the cluster membership of these blue BO galaxies, a number of groups (e.g. Dressler \& Gunn 1983; Couch \&
Sharples 1987) undertook spectroscopic surveys of distant clusters and came up with surprising results. Although
some blue galaxies are found to be interloping foreground or background galaxies, the {\it bona fide} blue cluster
members are found to exhibit spectra typical of spiral galaxies. They display strong emission lines indicative of
ongoing star formation.  Further, some red galaxies show strong Balmer absorption lines with a lack of emission
lines, indicative of a recent decline in star formation (Figure \ref{fig:p99}).

%

This latter type subsequently became known as post-starbursting galaxies (PSGs) or `E+A'\footnote{The nomenclature
`E+A' derives from the dominant presence of young, luminous, blue, A-type stars (resulting from a short-lived
episode of star formation) contained within a galaxy possessing an old stellar population (E). Note that Poggianti
et al.\ (1999) have further refined this class into `k+a' and `a+k' species.} galaxies (e.g.\ Franx 1993;
Zabludoff et al.\ 1996). The presence of the E+A galaxies in these clusters is surprising given their comparative
rarity in nearby clusters (Caldwell \& Rose 1997) and further reaffirms the view that clusters of galaxies have
been recently evolving.

A complicating issue to the BO effect is the morphology--density ($T-\Sigma$) relation first discussed by Dressler
(1980).  Table \ref{tab:TS} provides an illustration of the $T-\Sigma$ relation sufficient for this discussion:
the proportion of spiral galaxies decreases strongly as a function of local galaxy density whilst lenticular (S0)
and elliptical (E) galaxies increase. Oemler (1974) demonstrates a similar result within a number of relaxed, rich
clusters with the E and S0 fraction increasing towards the centre, accompanied by a decline in the fraction of
spiral types. The fraction of early types is also found to be correlated with global cluster structure: more
relaxed clusters have a higher early-type fraction (Oemler 1974). The question of whether local galaxy density or
clustocentric radius is the more fundamental parameter driving the morphological transformation, however, is still
an open question (Whitmore \& Gilmore 1991; Whitmore, Gilmore, \& Jones 1993).

Moreover, recent studies by Rakos, Odell, \& Shombert (1997) and Margoniner \& de Carvalho (2000) show that the
bluer galaxies are indeed preferentially located in low density environs. This is particularly valid in
observations of relaxed systems (e.g.\ Smail et al.\ 1998) where $f_B$ is computed to be smaller than that
predicted by Butcher \& Oemler (1984). In cases where the clusters sampled are unrelaxed or contain significant
substructure, however, $f_B$ is found to be systematically larger (Caldwell \& Rose 1997; Metevier, Romer, \&
Ulmer 2000).

%
\begin{table}
\begin{center}
\caption{\small{An illustration of the morphology--density relation sourced from Oemler (1992). Because a number
of sources were used in the construction of this table, the populations vary within each category and therefore
these numbers should be taken as \emph{representative} of the trend and not absolute. See also Figure 4 of
Dressler (1980). }}\smallskip
\begin{tabular}{lccc}
\hline
        & \multispan3{\hfil \% Type \hfil } \\
Environment  & E & S0 & Spiral \\ \hline
\noalign{\smallskip}
Field           &  10 & 20 & 70 \\
Poor Group      &  10 & 20 & 70 \\
Rich Group      &  10 & 30 & 60 \\
Cluster         &  20 & 40 & 40 \\
\hline
\end{tabular}
  \label{tab:TS}
\end{center}
\end{table}

\section{Discussion}

Much recent effort has gone into attempts to tie together and understand the effects outlined above.  In modern
times this has been facilitated with the advent of the Hubble Space Telescope (HST) and advances in computational
power that permit more accurate simulations of clusters. The HST has provided an unrivalled tool for examination
of galactic morphologies at high redshift.

Two groups attempted to analyse morphological information provided by the HST before its refurbishment (Couch et
al.\ 1994; Dressler et al.\ 1994).  Independently, they made preliminary connections between morphologically
disturbed-looking galaxies and blue BO galaxies together with an examination of numerous elliptical galaxies
possessing E+A qualities (Couch et al.\ 1994).

Following the refurbishment of the HST, the two groups combined their efforts in the MORPHS collaboration (Smail
et al.\ 1997). The MORPHS sample covers 11 fields in 10 clusters in the redshift range $0.37<z<0.56$. They
demonstrate that the fraction of elliptical galaxies within the rich clusters remains near constant whilst the S0
fraction decreases with increasing redshift: a redshift evolution in the $T-\Sigma$ relation (Dressler et al.\
1997). Meanwhile, the $T-\Sigma$ relation is generally found to be stronger in those clusters which are more
relaxed and centrally concentrated. They conclude that whilst the elliptical galaxy population would have predated
the gravitational collapse of a cluster, the S0 population is only generated in large numbers after this. The blue
BO galaxies, meanwhile, appear morphologically disturbed and observations of their emission lines indicate that
they are indeed currently star-forming (Couch et al.\ 1998).

\begin{figure}
\centerline{\psfig{file=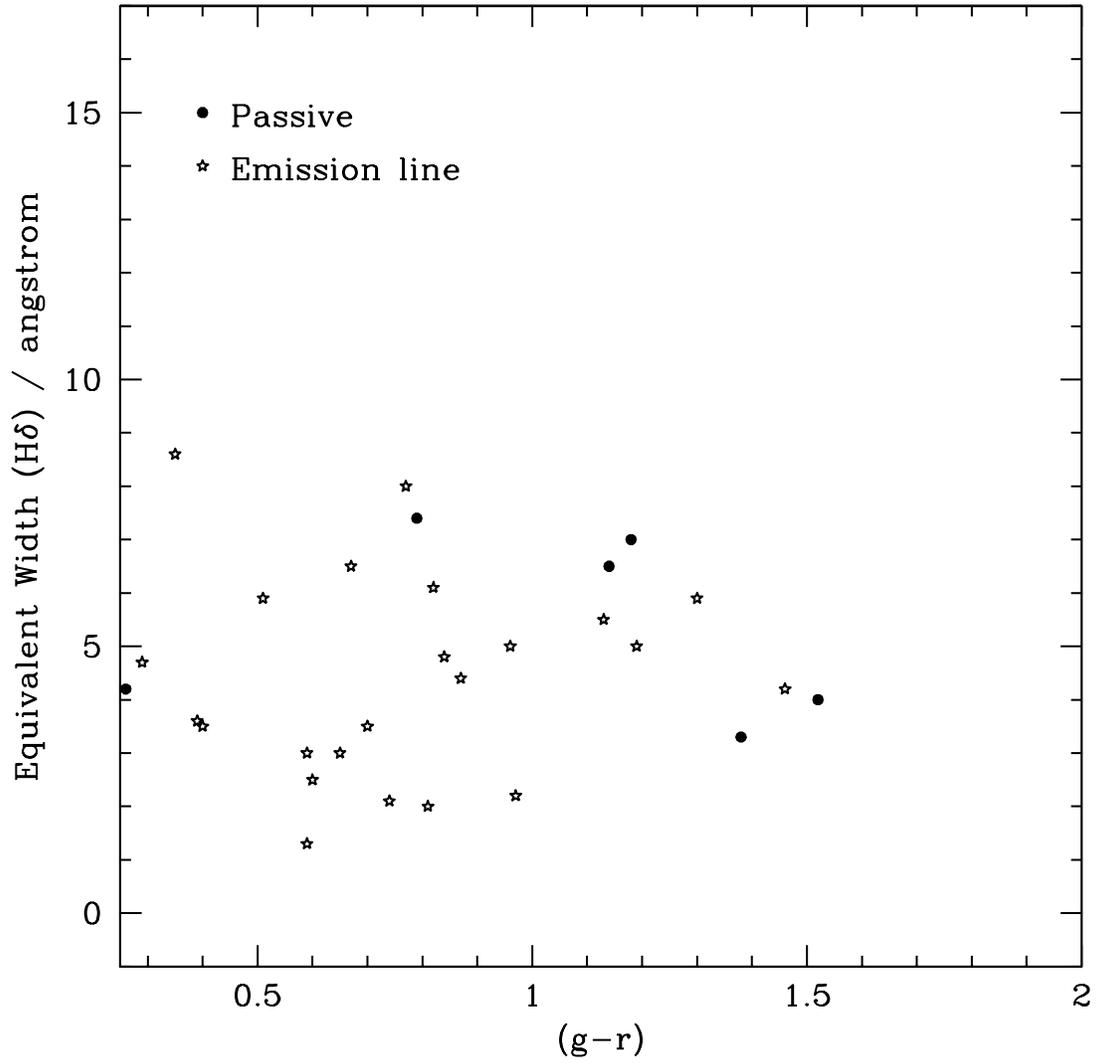,angle=0,width=6.in}}
  \caption[Data from Poggianti et al.\ (1999)]{\small{As
for Figure \ref{fig:p99} but for a field sample. The incidence of passive galaxies that have
experienced recent star formation (as inferred from H$\delta$ equivalent widths) is markedly
lower than for the cluster sample (Figure \ref{fig:p99}).}}
  \label{fig:p99f}
\end{figure}

More recent work by the same collaboration suggests that recently star-forming galaxies constitute $\sim 20$ per
cent of the total population (Dressler et al.\ 1999; Figure \ref{fig:p99}). In contrast, their incidence in the
field at the same redshift is found to be lower (two per cent; Figure \ref{fig:p99f}). Further, the frequency of
recently star-forming galaxies belonging to clusters at these redshifts is greater than equivalent clusters at
lower redshifts and has a stronger evolution with redshift than field E+As suggesting a strong evolution in this
galaxy type. Those galaxies found to be star-forming (i.e.\ possessing emission lines) also exhibit a larger
spatial extent and greater velocity dispersion than the passively evolving population. Correspondingly, the
recently star-forming galaxies exhibit qualities intermediate to the star-forming and passive populations
(Dressler et al.\ 1999; Poggianti et al.\ 1999).

%

Another major modern study of high redshift ($z=0.2-0.55$) clusters is that of the Canadian Network for
Observational Cosmology (CNOC: e.g. Yee, Ellingson, \& Carlberg 1996; Ellingson et al.\ 2001). The primary CNOC
sample is composed of 16 X-ray selected clusters of galaxies observed with the Canada--France--Hawai'i Telescope.
Over 6400 spectra have been obtained for this sample, of which over 2600 have provided a reliable redshift.

Fifteen of the CNOC clusters have been analysed by Balogh et al.\ (1999). In contrast to MORPHS, they find that
the fraction of star-forming galaxies only weakly increases with redshift. They further find that the recently
star-forming population does not increase with redshift with any significance and is not in excess to the value
for the local field.

The CNOC have also examined the dependence of galaxy type with clustocentric radius: they report an absence of
emission line galaxies within the cluster region correlating well with radius (Balogh et al. 1997).  This
correlation, however, cannot wholly be accounted for by the $T-\Sigma$ relation (Balogh et al.\ 1998). Several
other cluster gradients are also reported by Abraham et al.\ (1996) and Morris et al.\ (1998) including a
dependence of galaxy colour with clustocentric radius (i.e.\ environment). Work by Pimbblet et al.\ (2002) appears
to confirm this finding with higher redshift clusters displaying a stronger radial blueing of the CMR. This can be
interpreted as an age effect, whereby those galaxies at the outskirts of the cluster have younger stellar
populations than those in the core region.

How is it possible to reconcile the significantly lower fraction of star-forming and post star-forming galaxies
found by Balogh et al.\ (1997) than by Dressler et al.\ (1999), given that their cluster samples are coeval? A
number of possibilities are explored by Ellingson et al. (2000).  They attribute the discrepancies to relative
differences in cluster selection technique: CNOC used X-ray selected clusters\footnote{The CNOC clusters, being
X-ray selected, are mostly relaxed, regular clusters, many of which contain cooling flows indicative of a lack of
recent merging activity (Allen 1998).} whereas MORPHS used optical selection.\footnote{The MORPHS clusters were
collected over a number of years during the 1990s and are composed of both optically- and AGN-selected clusters
which span a range of richness and morphology.} The CNOC spectral data also possess a lower signal-to-noise ratio
than MORPHS and therefore may not be able to classify recently star-forming galaxies as accurately.

A potential complicating factor is the existence of `dusty-starburst galaxies' reported by Poggianti et al.\
(1999).  These galaxies exhibit both strong emission and absorption lines.  It has been suggested that these
potential PSG progenitors have had their star-formation rates underestimated (Poggianti et al.\ 1999) owing to
extinction.  The star-formation rate is estimated by the strength of the [OII] emission line.  This line, however,
is sensitive to dust extinction.  Some effort is now going into using less dust-sensitive emission lines as
indicators of star formation (e.g. H$\alpha$: Balogh \& Morris 2000).

At lower redshift ($z<0.1$) Zabludoff et al.\ (1996) use the Las Campanas Redshift Survey (LCRS; Shectman et al.\
1996) to find that recently star-forming (E+A) galaxies in the field constitute a low fraction ($0.2$ per cent) of
the total.  This is in agreement with the CNOC collaboration, yet is significantly smaller than that found by the
MORPHS collaboration in the field. Dressler et al.\ (1999) argued that the increase in recently star-forming
galaxies in the field is a factor of two between their two samples.

\section{Environmental Mechanisms}

Environmental conditions can provide mechanisms to explain the observations.  Assuming that initial condition
considerations result in all protogalaxies being the same ($\sim$ spiral), there are several processes that can
take place which have the potential to explain the observations.

One classic example is that of \emph{ram pressure stripping} (or ram pressure ablation: Gunn \& Gott 1972). On
impact with the intra-cluster medium, galaxies that are infalling onto the cluster have their gas removed. In the
case of a spiral galaxy, the stripping of its gaseous disc would result in the creation of a lenticular cluster
galaxy (Quilis, Moore, \& Bower 2000). Unfortunately, this mechanism does not explain the presence of S0 galaxies
in low density environments and certainly cannot be the sole cause for the observed variations in galaxy
properties with environment (Lewis et al.\ 2002; G{\' o}mez et al.\ 2003).

Galaxy--galaxy interactions could also provide a mechanism for the morphological transformation of spirals into
ellipticals or lenticulars (Toomre \& Toomre 1972; Barnes \& Hernquist 1991) and enhance star formation (Lavery \&
Henry 1988; Lavery, Pierce, \& McClure 1992; Rakos, Maindl, \& Schombert 1996). Certainly there are enough
strongly interacting galaxies observed to support this claim (Toomre 1978). Whilst this kind of interaction may
explain some of the blue BO galaxies there are certainly not enough galaxy--galaxy interactions to explain all of
them (Smail et al.\ 1997; Ghigna et al.\ 1998). Moore et al.\ (1996) propose a modified form of this interaction
called `harassment' whereby just close encounters of galaxy pairs is enough to initiate morphological
transmutation.

Tidal interactions may also be taking place between galaxies and the gravitational potential of the cluster itself
(Bekki, Couch, \& Shioya 2001a). This process should be a more effective mechanism than ram pressure stripping in
causing activity (Byrd \& Valtonen 1990; Rose et al.\ 2001).  Such a mechanism would, however, produce BO galaxies
near the cluster core (Fujita 1998) which is observationally inconsistent.

Morphological segregation could also be driven by a slow starvation of star-forming gas (Larson, Tinsley, \&
Caldwell 1980; Bekki, Couch, \& Shioya 2002). This could be caused by the removal of the gas reservoirs as
galaxies enter the cluster (Poggianti et al.\ 1999; Balogh, Navarro, \& Morris 2000). The mechanism may lead
directly to the formation of an S0 galaxy, or other processes may be involved (Balogh et al.\ 2001). Whatever the
exact details of the process that suppresses the star formation of infalling galaxies, it necessarily must be able
to operate in low density environments (Fasano et al.\ 2000; Bekki et al.\ 2001b; Carlberg et al.\ 2001; Couch et
al.\ 2001; Balogh et al.\ 2002; also see Drinkwater et al.\ 2003). One problem with the starvation mechanism is
the inherent gentility: it does not produce enough starbursting galaxies (Oemler 1992).

Observationally, whilst Poggianti et al.\ (1999) favour an abrupt truncation of the star-formation rate to explain
the frequency of recently star-forming galaxies, Balogh et al.\ (1999) advocate a more steady decline to explain
the radial dependence of emission line strength. Kodama \& Smail (2001) report that whatever the nature of the
morphological transformation mechanism, the process requires 1--3\,Gyr to complete from initial infall and that
clusters must necessarily have a high accretion rate.

\section{Selection Effects and Biases}

To be certain that the BO effect is real, one must account for the way in which cluster samples are selected. It
could be that the clusters observed at higher redshift are not comparable to lower redshift ones.  Oemler,
Dressler \& Butcher (1997) proposed that higher redshift clusters are more exceptional than their lower redshift
counterparts as they are dynamically younger and at an epoch of enhanced star formation.

\begin{figure}
\centerline{\psfig{file=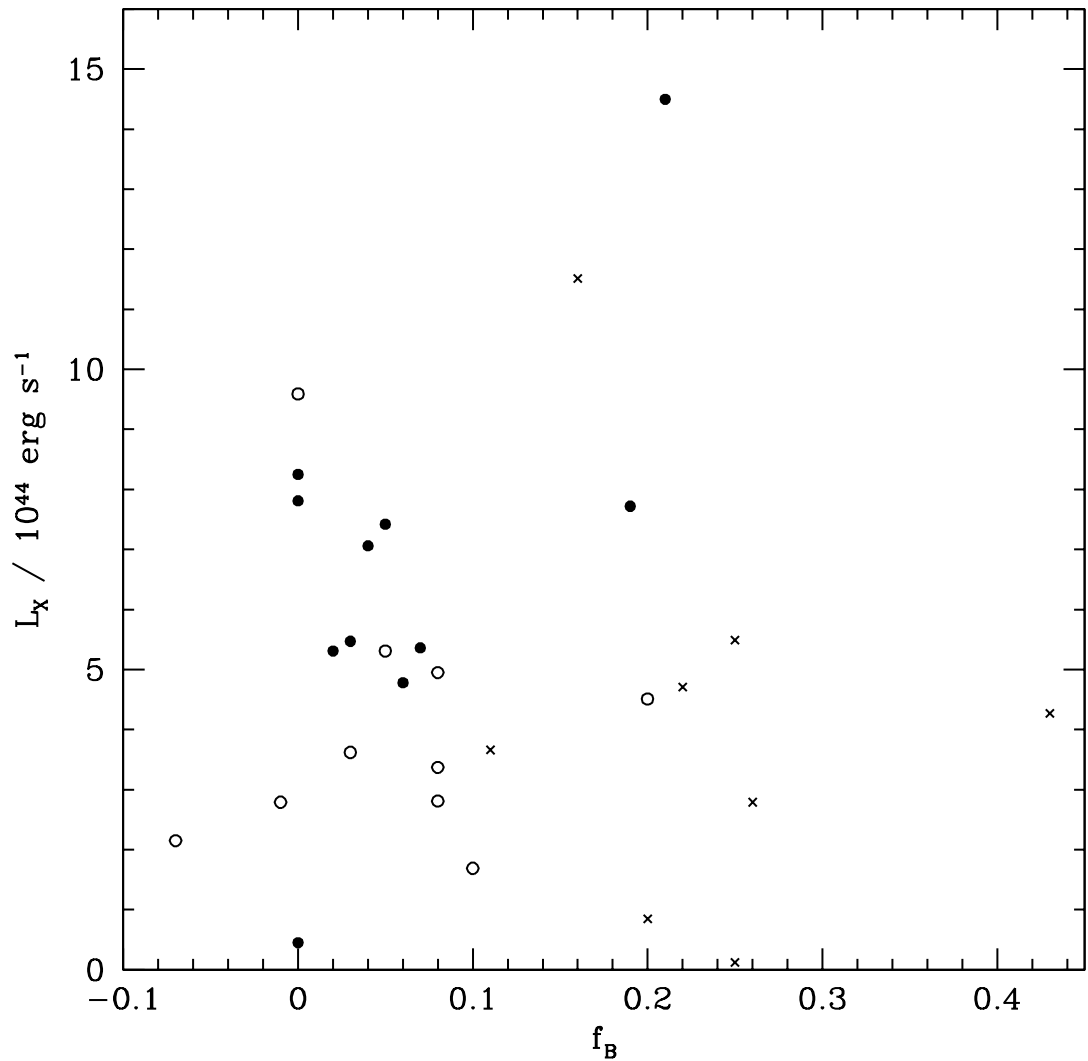,angle=0,width=6.in}}
  \caption[Blue fraction against X-ray luminosity]{\small{Blue fraction,
$f_B$, plotted as a function of $L_X$ from Pimbblet et al.\ (2002: filled circles), Fairley et al.\ (2002:
crosses), and Smail et al.\ (1998: open circles). Each set of points comes from a different redshift range ($0.08
< z < 0.15$, $0.24 < z < 0.58$, $0.22 < z < 0.28$, respectively). This is no significant correlation found between
these parameters. }}
  \label{fig:fblx}
\end{figure}

Andreon \& Ettori (1999) take this thought process further. They demonstrate that the mean X-ray luminosity
($L_X$) of the Butcher \& Oelmer (1984) clusters increases with redshift. Therefore, the redshift trend in $f_B$
might not be due to one object class.\footnote{Cluster X-rays are produced from thermal bremsstrahlung of the
(hot) inter-cluster gas (e.g.\ Jones \& Forman 1984; Sarazin 1988). Larger X-ray luminosity, therefore, is
indicative of more relaxed, massive clusters (Edge \& Stewart 1991; see also Zabludoff \& Zaritsky 1995 for a
perspective on X-rays from a merging cluster).} If true, this has implications for studies such as Ebeling, Edge
\& Henry (2001) who seek out only the most massive clusters at high redshifts. Figure \ref{fig:fblx} displays
$f_B$ as a function of $L_X$ from a selection of published work.  This shows no significant correlation between
the two parameters.

Margoniner et al.\ (2001) circumvent this issue by selecting an unprecedented 295 Abell clusters (Abell 1958;
Abell, Corwin, \& Olowin 1989) with no {\it a priori} bias to richness class, $L_X$ (and hence mass), or
morphological type (and/or degree of substructure) as some other studies do (Smail et al.\ 1998; Balogh et al.\
2002; Fairley et al.\ 2002; Pimbblet et al.\ 2002). The results of Margoniner et al.\ (2001) show that the
evolution in $f_B$ is real and occurs at similar rates for clusters of all richness classes, with poorer clusters
(groups) displaying higher $f_B$ values at any given redshift. Lastly,
Margoniner et al.\ (2001) re-affirm the view that there is no significant trend of $f_B$ with $L_X$ (see also
Figure \ref{fig:fblx}).

%

Calculation of $f_B$ crucially depends upon the parts of the luminosity function that are probed.  Whilst Butcher
\& Oemler (1984) use $M_V = -20$ as their fiducial magnitude cut, other investigators vary (e.g.\ Margoniner et
al.\ 2001 sample down to $M^* + 2$). Sampling fainter magnitude cluster galaxies produces two major effects.
Firstly, the total number of galaxies in a cluster will increase.  This is beneficial as it will make Poissonian
errors smaller.  Secondly, different types of galaxy are sampled.  This makes comparison to other works harder to
facilitate as it can seriously affect the value of $f_B$ because fainter galaxies tend to have different colours.

Choice of the physical radius analysed (e.g.\ $R_{30}$) can also vary $f_B$ (see Ellingson et al.\ 2001). This is
simple to understand in terms of the $T-\Sigma$ relation (Dressler 1980) as at larger radii, more blue, late-type
galaxies will be present and cluster substructure can become important. Interestingly, by converting their
selection criteria to that of the original BO work, Margoniner et al.\ (2001) find stronger cluster evolution than
Butcher \& Oemler did. This is consistent with the results of Rakos \& Schombert (1995) and Margoniner \& de
Carvalho (2000).

In studies without spectroscopic data, there will inevitably be interlopers from the field population that change
$f_B$ (e.g.\ Pimbblet et al. 2002).  Diaferio et al.\ (2001) investigated this contamination by making $N$-body
simulations.  Whilst interlopers may represent only a small fraction of the total cluster population, they find
that they are a significant fraction (perhaps up to 50 per cent) of the bluer BO population, particularly at small
clustocentric radii (i.e.\ within $R_{30}$).

\section{Concluding Remarks}

Finally, the results from two of the largest modern surveys, the 2dF Galaxy Redshift Survey (Lewis et al.\ 2002)
and the Sloan Digital Sky Survey (G{\' o}mez et al.\ 2003), are proving to be interesting. Both collaborations
agree that the decrease in star-formation rates of galaxies in cluster environments is a Universal phenomenon that
occurs over a wide range in density. The $T-\Sigma$ relation alone is unlikely to explain the origin of this
$\Sigma$ star-formation correlation (G{\' o}mez et al.\ 2003).

In conclusion, the landmark study of Butcher \& Oemler (1984) is merely the tip of an evolutionary iceberg
incorporating photometric, spectroscopic, and morphological changes. Whilst the debate remains intense, the
precise mechanism(s) required to drive these observations remains unknown. Future avenues of investigation should
concentrate on a multiwavelength (wide-field) approach tied with spectroscopy (to avoid contamination effects) and
must possess well defined selection criteria that can readily be compared to (or transformed directly into the
selection criteria of) previous studies. One line of inquiry would be to obtain data of clusters at high redshift
($z>0.5$ and beyond) to fill in the currently sparsely-sampled region of the $f_B$--$z$ plane.  Although
observationally challenging, the advent of 8\,m telescopes should make this task easier, although one must be
careful in the cluster selection. Other uninvestigated areas include how $f_B$ varies with cluster environment
within superclusters, and sampling reasonably large galaxy groups (similar to Balogh et al.\ 2002) to probe the
extrema of the mass scale. When tied in with X-ray observations, a clearer picture of the effect that substructure
plays should also become apparent.

\section*{Acknowledgments}

KAP thanks the many and varied discussions with the staff at
the University of Durham, plus the rest of the LARCS team.
The anonymous referee also provided a number of
constructive suggestions which have improved this work.
This work was supported from an EPSA University of Queensland
Research Fellowship.

\section*{References}

\reference Abell, G.O.\ 1958, \apjs,  3, 211

\reference Abell, G.O., Corwin, H.G., \& Olowin, R.P.\ 1989, \apjs,  70, 1

\reference Abraham, R.G., et al.\ 1996, \apj, 471, 694

\reference Allen, S.W.\ 1998, \mnras, 296, 392

\reference Andreon, S., \& Ettori, S.\ 1999, \apj,  516, 647

\reference Balogh, M.L., \& Morris, S.L.\ 2000, \mnras, 318, 703

\reference Balogh, M.L., Navarro, J.F., \& Morris, S.L.\ 2000, \apj, 540, 113

\reference Balogh, M.L., Christlein, D., Zabludoff, A.I., \& Zaritsky, D.\ 2001, \apj, 557, 117

\reference Balogh, M.L., Morris, S.L., Yee, H.K.C., Carlberg, R.G., \& Ellingson, E.\ 1997, \apjl, 488, L75

\reference Balogh, M.L., Morris, S.L., Yee, H.K.C., Carlberg, R.G., \& Ellingson, E.\ 1999, \apj, 527, 54

\reference Balogh, M.L., Schade, D., Morris, S.L., Yee, H.K.C., Carlberg, R.G., \& Ellingson, E.\ 1998, \apjl,
504, L75

\reference Balogh, M., Bower, R.G., Smail, I., Ziegler, B.L., Davies, R.L., Gaztelu, A., \& Fritz, A.\ 2002,
\mnras, 337, 256

\reference Barnes, J.E., \& Hernquist, L.E.\ 1991, \apjl, 370, L65

\reference Bekki, K., Couch, W.J., \& Shioya, Y.\ 2001a, \pasj,  53, 395

\reference Bekki, K., Couch, W.J., \& Shioya, Y.\ 2002, \apj,  577, 651

\reference Bekki, K., Couch, W.J., Drinkwater, M.J., \& Gregg, M.D.\ 2001b, \apjl, 557, L39

\reference Bower, R.G., Lucey, J.R., \& Ellis, R.S.\ 1992, \mnras, 254, 601

\reference Butcher, H., \& Oemler, A.\ 1978a, \apj, 219, 18

\reference Butcher, H., \& Oemler, A.\ 1978b, \apj, 226, 559

\reference Butcher, H., \& Oemler, A.\ 1984, \apj, 285, 426

\reference Butcher, H.R., \& Oemler, A.\ 1985, \apjs, 57, 665

\reference Butcher, H., Wells, D.C., \& Oemler, A.\ 1983, \apjs, 52, 183

\reference Byrd, G., \& Valtonen, M.\ 1990, \apj, 350, 89

\reference Caldwell, N., \& Rose, J.A.\ 1997, \aj, 113, 492

\reference  Carlberg, R.G., Yee, H.K.C., Morris, S.L., Lin, H., Hall, P.B., Patton, D.R., Sawicki, M., \&
Shepherd, C.W.\ 2001, \apj, 563, 736

\reference Couch, W.J., \& Newell, E.B.\ 1984, \apjs, 56, 143

\reference Couch, W.J., \& Sharples, R.M.\ 1987, \mnras, 229, 423

\reference Couch, W.J., Ellis, R.S., Sharples, R.M., \& Smail, I.\ 1994, \apj, 430, 121

\reference Couch, W.J., Barger, A.J., Smail, I., Ellis, R.S., \& Sharples, R.M.\ 1998, \apj, 497, 188

\reference Couch, W.J., Balogh, M.L., Bower, R.G., Smail, I., Glazebrook, K., \& Taylor, M.\ 2001, \apj, 549, 820

\reference Diaferio, A., Kauffmann, G., Balogh, M.L., White, S.D.M., Schade, D., \& Ellingson, E.\ 2001, \mnras,
323, 999

\reference Dressler, A.\ 1980, \apj, 236, 351

\reference Dressler, A., \& Gunn, J.E.\ 1983, \apj, 270, 7

\reference Dressler, A., Oemler, A.J., Butcher, H.R., \& Gunn, J.E.\ 1994, \apj, 430, 107

\reference Dressler, A., Smail, I., Poggianti, B.M., Butcher, H., Couch, W.J., Ellis, R.S., \& Oemler, A.J.\ 1999,
\apjs, 122, 51

\reference Dressler, A., et al.\ 1997, \apj, 490, 577

\reference Drinkwater, M.J., Gregg, M.D., Hilker, M., Bekki, K., Couch, W.J., Ferguson, H.C., Jones, J.B., \&
Phillipps, S.\ 2003, \nat, 423, 519

\reference Ebeling, H., Edge, A.C., \& Henry, J.P.\ 2001, \apj,  553, 668

\reference Edge, A.C., \& Stewart, G.C.\ 1991, \mnras,  252, 414

\reference Ellingson, E., Lin, H., Yee, H.K.C., \& Carlberg, R.G.\ 2001, \apj,  547, 609

\reference Ellingson, E., Abraham, R., Yee, H.K.C., Lin, H., Laurent, G., \& Schade, D.\ 2000,
Bulletin of the American Astronomical Society, 197, 57.01

\reference Fairley, B.W., Jones, L.R., Wake, D.A., Collins, C.A., Burke, D.J., Nichol, R.C., \& Romer, A.K.\ 2002,
\mnras,  330, 755

\reference Fasano, G., Poggianti, B.M., Couch, W.J., Bettoni, D., Kj{\ae}rgaard, P., \& Moles, M.\ 2000, \apj,
542, 673

\reference Franx, M.\ 1993, \apjl, 407, L5

\reference Fujita, Y.\ 1998, \apj, 509, 587

\reference Ghigna, S., Moore, B., Governato, F., Lake, G., Quinn, T., \& Stadel, J.\ 1998, \mnras, 300, 146

\reference G{\' o}mez, P.L., et al.\ 2003, \apj, 584, 210

\reference Gunn, J.E., \& Gott, J.R.I.\ 1972, \apj, 176, 1

\reference Jones, C., \& Forman, W.\ 1984, \apj,  276, 38

\reference Kodama, T., \& Smail, I.\ 2001, \mnras, 326, 637

\reference La Barbera, F., Merluzzi, P., Iovino, A., Massarotti, M., \& Busarello, G.\ 2003, A\&A,  399, 899

\reference Larson, R.B., Tinsley, B.M., \& Caldwell, C.N.\ 1980, \apj, 237, 692

\reference Lavery, R.J., \& Henry, J.P.\ 1988, \apj,  330, 596

\reference Lavery, R.J., Pierce, M.J., \& McClure, R.D.\ 1992, \aj,  104, 2067

\reference Lewis, I., et al.\ 2002, \mnras, 334, 673

\reference Margoniner, V.E., \& de Carvalho, R.R.\ 2000, \aj,  119, 1562

\reference Margoniner, V.E., de Carvalho, R.R., Gal, R.R., \& Djorgovski, S.G.\ 2001, \apj,  548, L143

\reference Metevier, A.J., Romer, A.K., \& Ulmer, M.P.\ 2000, \aj,  119, 1090

\reference Moore, B., Katz, N., Lake, G., Dressler, A., \& Oemler, A.\ 1996, \nat, 379, 613

\reference Morris, S.L., Hutchings, J.B., Carlberg, R.G., Yee, H.K.C., Ellingson, E., Balogh, M.L., Abraham, R.G.,
\& Smecker-Hane, T.A.\ 1998, \apj, 507, 84

\reference Oemler, A.J.\ 1974, \apj, 194, 1

\reference Oemler, A.\ 1992, in Clusters and Superclusters of Galaxies, NATO ASIC Proceedings 366, ed.\ A. Fabian
(Dordrecht: Kluwer Academic Publishers), 29

\reference Oemler, A., Dressler, A., \& Butcher, H.\ 1997, \apj, 474, 561

\reference Pimbblet, K.A.\ 2001, PhD Thesis, University of Durham

\reference Pimbblet, K.A., Smail, I., Kodama, T., Couch, W.J., Edge, A.C., Zabludoff, A.I., \& O'Hely, E.\ 2002,
\mnras, 331, 333

\reference Poggianti, B.M., Smail, I., Dressler, A., Couch, W.J., Barger, A.J., Butcher, H., Ellis, R.S., \&
Oemler, A.J.\ 1999, \apj, 518, 576

\reference Quilis, V., Moore, B., \& Bower, R.\ 2000, Science, 288, 1617

\reference Rakos, K.D., \& Schombert, J.M.\ 1995, \apj, 439, 47

\reference Rakos, K.D., Maindl, T.I., \& Schombert, J.M.\ 1996, \apj,  466, 122

\reference Rakos, K.D., Odell, A.P., \& Schombert, J.M.\ 1997, \apj,  490, 194

\reference Rose, J.A., Gaba, A.E., Caldwell, N., \& Chaboyer, B.\ 2001, \aj,  121, 793

\reference Sarazin, C.L.\ 1988, X-ray Emission from Clusters of Galaxies (Cambridge: Cambridge University Press)

\reference Shectman, S.A., Landy, S.D., Oemler, A., Tucker, D.L., Lin, H., Kirshner, R.P., \& Schechter, P.L.\
1996, \apj, 470, 172

\reference Smail, I., Edge, A.C., Ellis, R.S., \& Blandford, R.D.\ 1998, \mnras, 293, 124

\reference Smail, I., Dressler, A., Couch, W.J., Ellis, R.S., Oemler, A.J., Butcher, H., \& Sharples, R.M.\ 1997,
\apjs, 110, 213

\reference Terlevich, A.I., Caldwell, N., \& Bower, R.G.\ 2001, \mnras, 326, 1547

\reference Toomre, A.\ 1978, in Large Scale Structures in the Universe, IAU Symposium 79, ed.\ M.S.\ Longair \&
J.\ Einasto (Dordrecht: Reidel), 109

\reference Toomre, A., \& Toomre, J.\ 1972, \apj, 178, 623

\reference Visvanathan, N., \& Sandage, A.\ 1977, \apj, 216, 214

\reference Whitmore, B.C., \& Gilmore, D.M.\ 1991, \apj, 367, 64

\reference Whitmore, B.C., Gilmore, D.M., \& Jones, C.\ 1993, \apj, 407, 489

\reference Yee, H.K.C., Ellingson, E., \& Carlberg, R.G.\ 1996, \apjs, 102, 269

\reference Zabludoff, A.I., \& Zaritsky, D.\ 1995, \apj,  447, L21

\reference Zabludoff, A.I., Zaritsky, D., Lin, H., Tucker, D., Hashimoto, Y., Shectman, S.A., Oemler, A., \&
Kirshner, R.P.\ 1996, \apj, 466, 104

\end{document}